\documentclass[amsfonts,prb,aps]{revtex4}
\usepackage{amsmath}
\usepackage{amsfonts}
\usepackage{epsfig}
\usepackage{amssymb}
\usepackage{graphicx}

\begin{document}

\title{Heat Capacity as A Witness of Entanglement}

\author{Marcin Wie\'sniak}
\affiliation{Faculty of Physics, University of Vienna,
Boltzmanngasse 5, A-1090 Wien, Austria}\affiliation{Instytut Fizyki Teoretycznej i Astrofizyki Uniwersytet Gda\'nski, PL-80-952 Gda\'nsk, Poland}
\affiliation{Erwin Schr\"odinger International
Institute for Mathematical Physics, Boltzmanngasse 9, A-1090 Wien,
Austria}
\affiliation{Present Address: Department of Physics, National University of Singapore, 2 Science Drive 3, Singapore 117542, Singapore}
\affiliation{Present Address: Centre for Quantum Technologies, National University of Singapore, 3 Science Drive 2, Singapore 117543, Singapore}
%\footnotemark[1]
\author{Vlatko Vedral}\affiliation{The School of Physics and Astronomy, University of Leeds, Leeds,
LS2 9JT, United Kingdom}\affiliation{Present Address: Department of Physics, National University of Singapore, 2 Science Drive 3, Singapore 117542, Singapore}
\affiliation{Present Address: Centre for Quantum Technologies, National University of Singapore, 3 Science Drive 2, Singapore 117543, Singapore}
%\footnotemark[2]
\author{{\v C}aslav Brukner}
\affiliation{Faculty of Physics, University of Vienna,
Boltzmanngasse 5, A-1090 Wien, Austria}\affiliation{Institut f\"ur
Quantenoptik und Quanteninformation, \"Osterreichische Akademie der
Wissenschaften, Boltzmanngasse 3, A-1090 Wien, Austria}

\begin{abstract}

We demonstrate that the presence of entanglement in macroscopic
bodies (e.g. solids) in thermodynamical equilibrium could be
revealed by measuring heat-capacity. The idea is that if the system was in a
separable state, then for certain
Hamiltonians heat capacity would not tend asymptotically to zero as
the temperature approaches absolute zero. Since this would contradict the third law of
thermodynamics, one concludes that the system must contain
entanglement. The separable bounds are obtained by minimalization of the heat capacity over separable states and using its universal
low-temperature behavior. Our results
open up a possibility to use standard experimental
techniques of solid-state physics -- namely, heat capacity measurements -- to
detect entanglement in macroscopic samples.

\end{abstract}

\date{\today}

%\pacs{03.67.Hk, 03.65.Ta, 03.65.Ud}

\maketitle

\section{Introduction}

Entanglement is not only a fundamental and curious feature of purely
quantum nature~\cite{schroedinger}, but it is recognized
as a physical resource useful in tasks such as quantum computation,
quantum cryptography or reduction of communication complexity. A
state $\rho$ of $N$ subsystems is entangled if it cannot be prepared by local
operations and classical communication, i.e. it cannot be written as
a convex sum over product states: $\rho = \sum_{j} w_j
\rho^{(1)}_j\otimes... \otimes \rho^{(N)}_j$, where the factorizable
state $\rho^{(1)}_j\otimes... \otimes \rho^{(N)}_j$ of $N$
individual systems occurs with the weight $w_j \geq 0$ $(\sum_j
w_j=1$) in the mixture.

A particularly interesting question is whether or not microscopic phenomena, 
such as quantum correlations between individual constituents of a
macroscopic body (for example, individual spins of a magnetic solid)
may affect its macroscopic properties.  The usual expectation is
that non-classical effects in a macroscopic body vanish due to
interaction of its many degrees of freedom with the environment
(decoherence). In order to experimentally test this claim, one needs to apply some types
of entanglement criteria, such as the Bell inequalities~\cite{Bell}, to
macroscopic bodies. Due to limited access to the state of these
systems this is, however, usually not possible. Recently, it was
shown that some thermodynamical properties, such as the internal
energy~\cite{wang,qph0406040,toth,dowling,wu}, or the magnetic
susceptibility~\cite{bruknervedralzeil,wiesniaksusceptibility} can
detect entanglement between microscopic constituents of the solid.
They can hence be used as entanglement witnesses~\cite{witness}.
Their additional advantage is their extendibility, that is their
proportionality to the size of the sample. For this reason we do not
need to know how much of the material is studied, but express the
quantities as specific (molar, per site, etc.). However, the two
macroscopic quantities mentioned above also have some drawbacks as
entanglement witnesses. The magnetic susceptibility can be applied
only to magnetic systems and a specific class of Hamiltonians
(isotropic\cite{wiesniaksusceptibility}). Determining the
internal energy at a given temperature might be a complicated
experimental task.

In this paper we show how entanglement in macroscopic samples and in
thermodynamical equilibrium can be detected by measuring heat
capacity. Our method of entanglement detection is both simple and
generic. Unlike internal energy, measuring specific heat of
a solid is a well-established experimental routine in solid state
physics. Furthermore, heat capacity is a generic property of
materials and can thus also be measured on non-magnetic systems (in
contrast to magnetic susceptibility). 

In a more general context, our result shows a new link between two 
fundamental theories, quantum
mechanics and thermodynamics (see Refs. 11-13
for other interesting links between the two theories). It is related
to the Nernst's theorem, also known as the third law of
thermodynamics~\cite{nernst}. In the original version, the theorem
states that the entropy at the absolute zero temperature is
dependent only on the degeneracy of the ground state. Alternatively,
it can be expressed as a requirement of unattainability of the
absolute zero temperature in a finite number of
operations~\cite{footnote}. This requires the specific heat to tend
asymptotically to zero as the temperature approaches the absolute
zero. We will show, however, that, for certain Hamiltonians and
under the assumption that the system is in a separable state, one
obtains a non-vanishing value for heat capacity (separable bound) as
the temperature approaches the absolute zero. Since this contradicts
the third law of thermodynamics one concludes that the system must
be in an entangled state. The separable bound for heat capacity is
obtained in two different ways, by direct minimization of the value
of heat capacity over separable states (the explicit example
considered is the Ising model in a transverse magnetic field), and
by referring to the universal behavior of heat capacity close to
absolute zero. Using these methods we obtain the range of physical
parameters (critical temperature and strength of the magnetic field)
for which entanglement is present in various classes of systems.

One might question the relevance of our method for entanglement
detection since the method requires knowledge of the Hamiltonian of
the system, and thus one could directly determine its eigenstates,
build thermal states therefrom and check their separability. As we
know, computation of eigenstates is in general a hard problem and
the origin of some major difficulties in solid state
physics. Furthermore, even if the thermal (mixed) state is known, it
is in general hard to find out whether it is separable or not. Our
method requires only knowledge of eigenvalues (partition function)
and can easily be experimentally implemented.

Consider a system descried by a Hamiltonian $H$ to be in a
thermodynamical equilibrium at a given temperature $T$. Its thermal
state is given by $\rho_T=\exp\left(-\frac{H}{kT}\right)/Z$, where
$k$ is the Boltzmann constant and
$Z=\mbox{Tr}[\exp\left(-\frac{H}{kT}\right)]$ is the partition
function. The knowledge of the partition function allows us to derive
all thermodynamical quantities. For example, the internal energy is
given by $U=-\partial\ln Z/\partial\beta=\mbox{Tr}[ H
\exp\left(-\frac{H}{kT}\right)]/Z=\mbox{Tr}[\rho_T H]$, where
$\beta=1/kT$. Similarly, the heat capacity, $C=\partial U/\partial
T=1/(kT^2)\partial^2 \ln Z/\partial\beta^2$, is proportional to the
variance of the Hamiltonian,
\begin{equation}
C=\frac{\Delta^2(H)}{kT^2}=\frac{1}{kT^2}(\langle H^2\rangle-\langle
H\rangle^2).
\end{equation}

We first consider the particular case of an Ising chain in a
transverse magnetic field and then consider a more general case.

\section{Transverse Ising model}

\subsection{Introduction}

The Hamiltonian of an Ising ring of $N$ spins-$\frac{1}{2}$ in a
transverse magnetic field is given by
\begin{equation}
H_{Ising}=J\sum_{i=1}^N\sigma_i^z\sigma_{i+1}^z+B\sum_{i=1}^N\sigma_i^x,
\end{equation}
where we assume the periodic boundary conditions $N+i\equiv i$. Here $B$ is the external transverse magnetic field, and $J$ denotes the coupling constant, taken in this section to be $1$. The Pauli matrices
$\sigma_i^j$ $(j=x,y,z)$ have the following actions on $i$th qubit:
$\sigma^z_i|k_i\rangle=(-1)^{k_i}|k_i\rangle$ and
$\sigma^x_i|k_i\rangle=|k_i\oplus 1\rangle$ $(k=0,1)$, and $\oplus$
denotes the summation modulo 2.

In the following we will prove that no product state belongs to the
eigenbasis of $H_{Ising}$. Therefore, the variance of $H_{Ising}$
and consequently heat capacity, cannot vanish within the set of
these states. Since taking a convex sum over product states can
only increase the variance, we will conclude that heat capacity
cannot vanish for the set of all separable states.

\subsection{Proof of non-separability of the eigenstates}

The proof that all eigenstates of the Hamlitonian $H_{Ising}$ are
entangled is by {\it reductio ad absurdum}. We first assume the
opposite, i.e. that at least one of the eigenstates is a product
state, and then arrive at a contradiction. Suppose that the product
state $|\psi\rangle=\otimes_{i=1}^N(a_i|0_i\rangle+b_i|1_i\rangle)$
is an eigenstate that is associated to the eigenvalue $E$ and
$a_i\neq 0$ and $b_i\neq 0$ for all $i$ (If this was not the case,
i.e. $a_i=0$ or $b_i= 0$ for some $i$, $|\psi\rangle$ would not be
an eigenstate, because the magnetic field term in $H_{Ising}$ flips
the state $|0\rangle$ to $|1\rangle$ and vice versa for every
qubit). Under the assumption that $|\psi\rangle$ was an eigenstate
with an eigenvalue $E$, the following would need to hold:
\begin{equation}
\label{proportional}
E=\frac{\langle 00...0|H_{Ising}|\psi\rangle}{\langle 00...0|\psi\rangle}=\frac{\langle 10...0|H_{Ising}|\psi\rangle}{\langle 10...0|\psi\rangle}.
\end{equation}
The proof is as follows. The two denominators are equal to
$\prod_{i=1}^{N}a_i$ and $\frac{b_1}{a_1}\prod_{i=1}^{N}a_i$. Under
the action of $H_{Ising}$ onto $|\psi\rangle$ in the expression on
the left, only the terms for which either all spins are in
the state $|0\rangle$ or for which one spin is in the $|1\rangle$
state and the rest in $|0\rangle$ remain. The numerator on the left is
hence equal to
$\prod_{i=1}^Na_i\left(N+B\sum_{i=1}^N\frac{b_i}{a_i}\right)$. Under
the action of $H_{Ising}$ onto $|\psi\rangle$ in the expression on
the right only the following terms give a contribution:
$|10...0\rangle$ (equal to $N-4$), $|0...0\rangle$ and the states in
which the first and one other spin are anti-aligned to the rest. The
second numerator results in
$\prod_{i=1}^Na_i\left\{\frac{b_1}{a_1}(N-4)+B\left[1+\frac{b_1}{a_1}\left(\sum_{i=1}^N\frac{b_i}{a_i}-\frac{b_1}{a_1}\right]\right)\right\}$.
After simplifying both sides of
Eq.~(\ref{proportional}), it reduces to a quadratic equation
\begin{equation}
\label{quadratic}
-B\left(\frac{b_1}{a_1}\right)^2-4\frac{b_1}{a_1}+B=0.
\end{equation}
Note that Eq. (\ref{quadratic}) has two solutions, some $\frac{b_1}{a_1}=x_0$ and $-\frac{1}{x_0}$. Interchanging all bras $\langle 0|$ and $\langle 1|$ in Eq.~(\ref{proportional}) we arrive to the same form of the equation as Eq. (\ref{quadratic}), but for the inverse of the fraction, $\frac{b_1}{a_1}$. This means that also $-x_0$ and $\frac{1}{x_0}$ must satisfy the equation. This would be true only if $x_0=\pm 1$, which is, however, not possible. QED. 

\subsection{Minimization of the variance of Hamiltonian over separable states}

We minimize the variance $\Delta^2(H_{Ising})$ over all separable
states. To find the minimal value of the variance for all separable
states it is sufficient to perform minimization over pure product
states only. As shown by Hofmann and Takeuchi
\cite{hofmanntakeuchi}, any convex mixture over product states
$\rho=\sum_i w_i|\phi_i\rangle\langle\phi_i|$ $(w_i\geq 0,
\sum_iw_i=1$ are weights of the product states in the mixture) can
only increase the variance of an observable $A$, with respect to the
variances $\Delta^2(A)_i$ of $A$ for individual states
$|\phi_i\rangle$:
%\begin{widetext}
\begin{eqnarray}
\label{htmixed}
\Delta^2(A)=\sum_{i}w_i\langle(A-\langle A\rangle)^2\rangle_i\nonumber=\sum_iw_i\left(\underbrace{\langle A^2\rangle_i-\langle A\rangle^2_i}_{=\Delta^2(A)_i}+\underbrace{(\langle A\rangle_i-\langle A\rangle)^2}_{\geq 0}\right)\nonumber\geq \sum_i w_i\Delta^2(A)_i,
\end{eqnarray}
with $\langle A\rangle_i=\langle\phi_i|A|\phi_i\rangle$. This
implies that the bound that is obtained for pure product states will
also be the bound for all separable (in general, mixed) states.

It can be shown that if the state is a product state, then one has
%\begin{widetext}
\begin{eqnarray}
\Delta^2(H_{Ising})=\left(N-\sum_{i=1}^N\langle\sigma^z_i\rangle^2\langle\sigma^z_{i+1}\rangle^2
+2\sum_{i=1}^N\langle\sigma^z_i\rangle\langle\sigma^z_{i+2}\rangle
-\sum_{i=1}^N\langle\sigma^z_i\rangle\langle\sigma^z_{i+1}\rangle^2\langle\sigma^z_{i+2}\rangle\right)\nonumber\\
-2B\sum_{i=1}^N\langle\sigma^z_i\rangle\langle\sigma^z_{i+1}\rangle(\langle\sigma^x_{i}\rangle+\langle\sigma^x_{i+1}\rangle)
+B^2\left(N-\sum_{i=1}^N\langle\sigma^x_i\rangle\right).
\label{ising-variance}
\end{eqnarray}
Due to the translational symmetry of the system we expect that the
product state, which minimizes this expression, is
(quasi-)translation invariant. We first consider product states with
a period of two sites. In such a state every odd spin is in the same
state $|\psi_1\rangle$ and, similarly, every even spin is in the
state $|\psi_2\rangle$. Therefore, the state of $N$ spins ($N$ is
here taken even) is
$|\psi\rangle=\left(|\psi_1\rangle|\psi_2\rangle\right)^{\otimes
N/2}$ and herein called two-translation invariant. The assumed
two-translation invariance of the state allows neighboring spins to
have anti-parallel $z$ components -- to minimize the interaction
energy -- and at the same time to have the $x$ components, all
anti-aligned to the magnetic field. Since in
Eq.~(\ref{ising-variance}) one has terms dependent on the
correlations between non-neighboring spins, one could expect that
the variance should be minimized over states with a period of more
than two sites. We have considered four-translation invariant states,
$|\psi\rangle=\left(|\psi_1\rangle|\psi_2\rangle|\psi_3\rangle|\psi_4\rangle\right)^{\otimes
N/4}$ (thus $N$ is divisible by 4) and have shown numerically, that
the same bound for the variance is obtained as in the case of
two-translational invariant states. We assume that the Bloch vectors
of $|\psi_1\rangle$ and $|\psi_2\rangle$ lie in the $xz$-plane, that
is $\forall_i$ $\langle\psi_i|\sigma^y|\psi_i\rangle=0$. Under these
assumptions the variance takes a form of
\begin{eqnarray}
\Delta^2(H_{Ising})&=&N\left(1+z_1^2+z_2^2-3z^2_1z_2^2\right)\nonumber\\
&-&2BN[z_1z_2(x_1+x_2)]\nonumber\\
&+&B^2\frac{N}{2}\left(2-x_1^2-x_2^2\right), \label{variance2}
\end{eqnarray}
where we have adopted the notation
$x_i=\langle\psi_i|\sigma^x|\psi_i\rangle=\sin\theta_i$ and
$z_i=\langle\psi_i|\sigma^z|\psi_i\rangle=\cos\theta_i$. Since the
expression (\ref{variance2}) is proportional to $N$, it is
convenient to discuss the specific heat capacity, that is the heat
capacity per spin.

\subsection{Results}

Figure 1 presents the results of a numerical minimization of
Eq.~(\ref{variance2}) over $\theta_1$ and $\theta_2$ as a function
of the magnetic field with units of $J=k= 1$. The plot
confirms that no eigenstate of the Hamiltonian has the proposed form
for any $B\neq 0$. In the limit of a very strong field, all spins
tend to orient themselves toward the field and build a product state, however, the interaction term
contributes the variance with a constant magnitude. This explains
why the curve in Fig. 1 does not tend to
zero as $B$ increases to infinity, but saturates at $B\approx 3.5$.

\begin{figure}
\centering
\includegraphics[width=7cm]{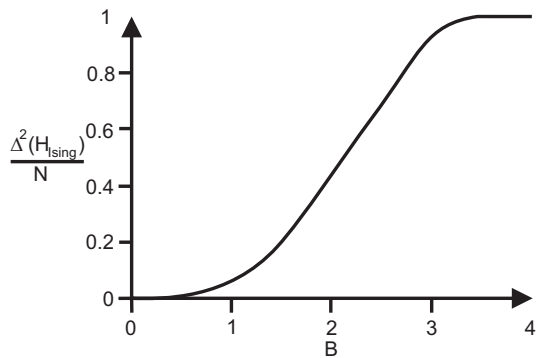}
\caption{The minimal variance per site $\Delta^2(H_{Ising})/N$ of
the Hamiltonian of a transverse Ising ring versus
the magnetic field $B$ ($J=k=1$). The minimization of the variance is performed over 2-translation
invariant product states
$\left(|\phi_{1}\rangle|\phi_{2}\rangle\right)^{\otimes N/2}$.} \label{gc_mu}
\end{figure}

\begin{figure}[t]
\centering
\includegraphics[width=8cm]{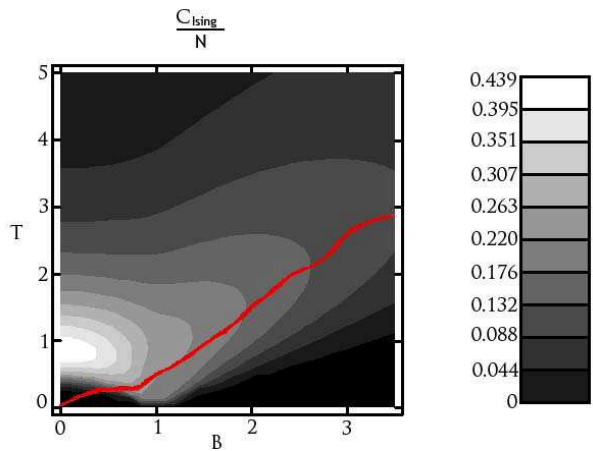}
\caption{(Color online) The specific heat per spin (in units $|J|=k=1$) of a
transverse antiferromagnetic Ising chain versus the temperature $T$
and the magnetic field $B$\cite{katsura}. The values
below the red line indicate the region of the $(T,B)$-space, where entanglement exists. There the variance of the Hamiltonian is lower than for any separable state.}
\end{figure}

The results of the optimization are compared to the values for
specific heat of an infinite Ising ring\cite{katsura}. Katsura\cite{katsura} obtained analytical forms of
thermodynamical quantities by an exact solution of the Hamiltonian
eigenproblem\cite{katsura}. The solution was achieved with the Jordan-Wigner
transformation followed by the Fourier and the Bogoliubov
transformations. The heat capacity per spin was found
to be
\begin{equation}
\label{heatkatsura} \frac{C_{Ising}}{N}=\frac{1}{\pi T^2}\int_0^\pi
\frac{f(B,\omega)}{\cosh^2\frac{f(B,w)}{T}}d\omega,
\end{equation}
where the Boltzmann constant is $k=1$ and $f(B,\omega)=\sqrt{1-2B\cos\omega+B^2}$. Figure 2
presents the expression~(\ref{heatkatsura}) for heat capacity per spin
as a function of the external magnetic field and temperature. Values
below the red line cannot be explained without entanglement; in this
region the $\Delta^2(H_{Ising})$ is lower than for any separable
state. An interesting observation is that with increasing the
strength of the magnetic field, the critical temperature below which
entanglement is detected increases as well.

Figure 3 presents the heat capacity per spin for a given magnetic
field $B=2$. The gray line represents $a/T^2$, where $a=0.4197$ is
the minimal Hamiltonian variance over separable states for this value of the field. For the temperatures below the value of intersection of the two lines the state is entangled.

\begin{figure}
\centering
\includegraphics[width=7cm]{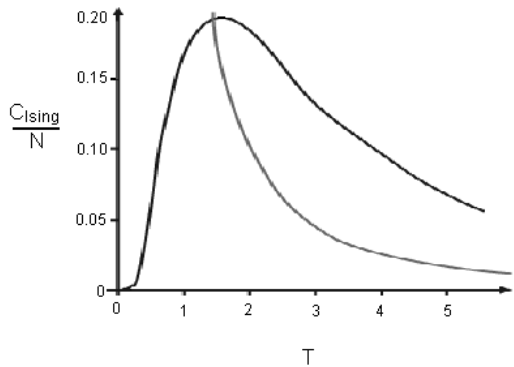}
\caption{The specific heat per spin (in
units $|J|=k=1$) of a transverse antiferromagnetic Ising chain versus
the temperature $T$ for $B=2$. The gray line, $0.4197/T^2$, is
our entanglement witness.}
\end{figure}

\section{Universal low-temperature behavior and entanglement}

In the following we will show that the specific heat is an
entanglement witness, whenever the internal energy is. The argument
is based on two results of statistical mechanics, universality and
scalability, which together state that the partition functions of
various systems are similar and characterized by a small number of
universal parameters, such as the central charge or the dimensionality
of a lattice.

\subsection{Internal energy as entanglement witness}

For the sake of the further discussion, we recall that for certain
Hamiltonians internal energy is an entanglement witness. This is nicely
illustrated in the case of the Heisenberg antiferromagnetic ring
Hamiltonian~\cite{qph0406040,toth}, given by
$H_{xxx}=J\sum_{i=1}^{N}\vec{s}^i\cdot\vec{s}^{i+1}$ (with
$\vec{s}^i$ denoting the $i$-th spin vector of the magnitude $s$ in a
ring, periodicity guaranteed by $i\equiv i+N$, and $J>0$). Under optimization over product states
$|\psi\rangle=\bigotimes_{i=1}^N|\psi_i\rangle$, the lowest possible
energy is $E_B=-JNs^2$, since for product states one has:
\begin{eqnarray}
E&=&\langle\psi|H_{xxx}|\psi\rangle=J\sum_{i=1}^{N}\langle\vec{s}^i\rangle\langle\vec{s}^{i+1}\rangle\nonumber\\
&\geq &-J\sum_{i=1}^{N}\sqrt{\langle\vec{s}^i\rangle^2}\sqrt{\langle\vec{s}^{i+1}\rangle^2}\geq-JNs^2\equiv E_B.
\label{bound}
\end{eqnarray}
Here, $\langle \vec{s}^{\! i}\rangle=\langle\psi_i|\vec{s}^i|\psi_i\rangle$.

By convexity, the same bound holds for all separable states. On the
other hand, in the limit of large $N$, the ground-state energy per
spin, for example, for $s=\frac{1}{2}$ is equal to -0.443$J$
\cite{hulthen}. Thus, in this case for all temperatures below a
critical one ($T_C$), for which the internal energy per spin is
$U_{xxx,T_C}=-0.25 J$, the state must contain entanglement.

\subsection{Gapless systems}

We demonstrate, how the separable bound on the internal energy can
be used to derive the separable bound on the heat capacity. First,
let us consider a class of materials, in which the lowest part of
the energetic spectrum is continuous. Infinite half-odd integer spin
chains and ring are examples of such systems. For sufficiently low
temperatures, their internal energy (hereafter, ``per spin"') can be
expanded to a polynome: $U(T)=E_0+(kT)^\gamma(c_0+c_1kT+...)$, with
$E_0$ being the ground-state energy, and $\gamma, c_0, c_1,...$ are
material-dependent constants. At sufficiently low temperatures, when
the higher order terms are negligible, the specific heat is
proportional to a power of $T$ as given by:
\begin{equation}
C(T)=\frac{\partial U(T)}{\partial T}=\gamma c_0k^\gamma
T^{\gamma-1}=\gamma\frac{U-E_0}{T}. \label{main1}
\end{equation}

Now we use the fact that if the internal energy is bounded for all
separable states, i.e., has a separable bound $E_B$, then the
specific heat is bounded for all separable states as well. Namely, one has
\begin{equation}
C(T)\geq\gamma\frac{E_B-E_0}{T}. \label{main2}
\end{equation}
This allows us to use the heat capacity as
entanglement witness. Given a Hamiltonian one determines the ground-state energy $E_0$ and the
minimal energy over all separable states, $E_B$. If the ground state is separable, i.e. $E_0=E_B$, the inequality
(\ref{main2}) is in agreement with the third law of thermodynamics,
as one can have $C\rightarrow 0$ with $T\rightarrow 0$. In the case
$E_0 \neq E_B$, one compares
$\gamma\frac{E_B-E_0}{T}$ with the real temperature dependence of
the heat capacity $C(T)$, obtained either from an experiment or
from a theory. Since $\gamma\frac{E_B-E_0}{T}$ diverges as the temperature
approaches the absolute zero and $C(T)$ has to tend to zero to be in an agreement
with the third law of thermodynamics, the two curves intersect each other.
The intersection point then defines the critical temperature below which entanglement exists in the system.
The method does not require the knowledge of the energy
eigenstates, but only of the ground state energy and a separable bound for
the internal energy.

We note that another thermodynamical entanglement witness has similar 
features to heat capacity. Namely, the magnetic susceptibity also might diverge in the
limit of infinitely low temperatures for all separable
states~\cite{wiesniaksusceptibility}. In both cases quantum
entanglement is decisive for finiteness of the low-temperature
values of the thermodynamical quantities.

Consider gapless models in 1+1 dimensions. For these systems, the
conformal field theory \cite{Affleck,blute,korepin} predicts $C=(\pi
c k^2T)/(3\hbar v)$ for the low-temperature behavior of the specific
heat (throughout this paragraph, we assume $J=k=\hbar=1$). Here $c$
stands for the central charge of the corresponding Virasoro algebra
and $v$ is the spin velocity. For half-integer spins $c=6s/(2+2s)$.
An example is the infinite xxx antiferromagnetic spin-$\frac{1}{2}$
Heisenberg chain. From inequality (\ref{main2}) one obtains that the
values of the specific heat below $C<0.386/T$ manifest entanglement
in the Heisenberg chain. The approximation $C\approx (2/3)T$ is
valid for temperatures below 0.1 \cite{xiang}.

Another example is the xx spin-$\frac{1}{2}$ antiferromagnets whose solution 
is also given in Ref. 17. The internal energy per spin of an infinite xx
ferromagnet described by the Hamiltonian
$H_{xx}=J\sum_i(\sigma_x^i\sigma_x^{i+1}+\sigma_y^i\sigma_y^{i+1})$
is given by
\begin{equation}
U_{xx}=-\frac{4J}{\pi}\int_0^{\pi/2}\tanh\left(\frac{2J}{kT}\sin\omega\right)\sin\omega d\omega.
\end{equation}
We will use the following approximation: at sufficiently low
temperatures $T<<k/J$, the argument $\tanh[(2Jx)/(kT)]$ of the
integral is a very steep function, which is well approximated  by
$f(x)=\left(\frac{2J}{kT}\right)x-\left(\frac{J}{kT}\right)^2x^2$
for $0\leq x\leq\frac{kT}{J}$ and $f(x)=1$ for $x>\frac{kT}{J}$.
Thus the internal energy can be written as
\begin{eqnarray}
\label{uxx}
U_{xx}&\approx&-\frac{4J}{\pi}\int_{0}^{\frac{kT}{J}}f(\omega)\omega d\omega-\frac{4J}{\pi}\int_{\frac{kT}{J}}^{\frac{\pi}{2}}\sin\omega d\omega\nonumber\\
&\approx&-\frac{5k^2T^2}{3\pi J}
-\frac{4J}{\pi}\left(1-\frac{k^2T^2}{2J^2}\right)\nonumber\\
&=&-\frac{4J}{\pi}+\frac{k^2T^2}{3\pi J},
\end{eqnarray}
where we have used expansions $\sin x\approx x$ and $\cos x\approx
1-x^2/2$. From Eq.~(\ref{uxx}) it is clear that the ground-state energy
per spin is $-4J/\pi$, while the energy bound for separable states
is $-J$. 
Hence, by Eq.~(\ref{main2})
the thermal state is entangled if it satisfies $C<2J(4/\pi-1)/T$ as
long as the specific heat can be approximated by a linear function
of $T$.

\subsection{Gapped systems}
The integer spin chains have an energy gap $\Delta$ between the
ground state and the first excited state, even in the
thermodynamical limit. The gap is also a feature of systems described by a Hilbert space of finite dimensions. 
For all gapped systems, the
low-temperature behavior of the specific heat is given by
\begin{eqnarray}
\label{main3}
C=c'T^\delta e^{-\frac{\Delta}{kT}},
\end{eqnarray}
where $c'$ and $\delta$ are some material-dependent constants
\cite{gap}. The internal energy can then be written as
$U(T)=U_0+\int_0^TC(T')dT'=U_0+c'(\Delta/k)^{\delta+1}\Gamma\left[-\delta-1,\Delta/(kT)\right	]$.
Since for large $x$ we have $\Gamma(a,x)\approx e^{-x}x^{a-1}$
\cite{Stegun}, we obtain $U(T)\approx
E_0+c'kT^{\delta+2}e^{-\Delta/(kT)}/\Delta$ at sufficiently low
temperatures. Finally, using the energy bound $E_B$ for separable
states we obtain the corresponding bound for the specific heat:
\begin{eqnarray}
\label{main4}
C\geq\frac{\Delta(E_B-E_0)}{kT^2}
\end{eqnarray}
for all separable states. Again, one can invoke the third law of
thermodynamics to argue for the necessity of existence of
entanglement at sufficiently low temperatures whenever $E_B\neq
E_0$. An exemplary gapped system is an infinite xxx spin-1
Heisenberg chain with the values $c'=\Delta^{5/2}/\sqrt{2\pi}$,
$\delta=-3/2$, $\Delta=0.411J$, and $E_0=-1.401J$
\cite{ref32,ref33}. By Eq. (\ref{bound}), $E_B=-J$. Thus, within the
approximation range, all low-temperature values of the specific heat
below $0.165J/(kT^2)$ cannot be explained without entanglement.

\section{conclusions}
In summary, we have shown that the low-temperature behavior of the
specific heat can reveal the presence of  entanglement in bulk bodies in the
thermodynamical equilibrium. For certain Hamiltonians and under the
assumption of having separable states only, the specific heat would diverge
at temperature approaching the absolute zero. This might be because none of
the eigenstates of the Hamiltonian are a product state and hence its
variance cannot vanish within the set of these states or, more
generally, because at least the ground state is entangled. In the latter
case we involve the separability bound and the universal low-temperature behavior of internal
energy to argue for non-classicality of a
thermal mixture in certain systems. One may therefore say that in these systems the 
validity of the third law of thermodynamics relies on quantum entanglement.

Thermal entanglement of bulk solids might play an important role in
the emerging quantum information technology, where non--classical
correlations were recognized as one of its main
resources~\cite{nielsen}. Our method enables to detect entanglement
using one of the standard techniques in solid-state physics --
measurements of the heat capacity.

\section{Acknowledgements}
We thank J. Kofler for valuable remarks. {\v C}.B. acknowledges
support of the Austrian Science Foundation (FWF), Doctorial Programe
"Complex Quantum Systems (QoCuS), and the European Commission (QAP).
M.W. is supported by the Erwin Schr\"odinger Institute in Vienna and
the Foundation for Polish Science (FNP) (including scholarship
START). This work is partially supported by the National Research Foundation and 
Ministry of Education, Singapore.

\end{document}